\documentclass[prl,aps,epsf,twocolumn,showpacs,10pt,superscriptaddress
]{revtex4-1}
\usepackage{graphicx,amsfonts,times,bm,amsmath,verbatim,color,array}

\usepackage{natbib}
\usepackage{graphicx,amssymb,amsmath,ifthen,braket,xcolor}
\usepackage{bm}

\usepackage{xr}
\makeatletter
\newcommand*{\addFileDependency}[1]{
  \typeout{(#1)}
  \@addtofilelist{#1}
  \IfFileExists{#1}{}{\typeout{No file #1.}}
}
\makeatother

\newcommand*{\myexternaldocument}[1]{
    \externaldocument{#1}
    \addFileDependency{#1.tex}
    \addFileDependency{#1.aux}
}

\myexternaldocument{SM}

\begin{document}

\title{Majorana corner modes and higher order topological superfluidity in an attractive Hubbard-Hofstadter model of cold atom optical lattices}

\author{Chuanchang Zeng}
\affiliation{Department of Physics and Astronomy, Clemson University, Clemson, South Carolina 29634, USA}

\author{T. D. Stanescu}
\affiliation{Department of Physics and Astronomy, West Virginia University, Morgantown, WV 26506, USA}

\author{Chuanwei Zhang}
\affiliation{Department of Physics, The University of Texas at Dallas, Richardson, Texas 75080, USA}

\author{V.W. Scarola}
\affiliation{Department of Physics, Virginia Tech, Blacksburg, Virginia 24061, USA}

\author{Sumanta Tewari}
\affiliation{Department of Physics and Astronomy, Clemson University, Clemson, South Carolina 29634, USA}

\begin{abstract}
Higher order topological superconductors hosting Majorana-Kramers pairs (MKPs) as corner modes have recently been proposed in a two-dimensional (2D) quantum spin Hall insulator (QSHI) proximity-coupled to unconventional cuprate or iron-based superconductors. Here, we show such MKPs can be realized using a conventional $s$-wave superfluid with a soliton in cold atom systems governed by the Hubbard-Hofstadter model. The  MKPs emerge in the presence of interaction at the corners defined by the intersections of line solitons and the one-dimensional edges of the system. Our scheme is based on the recently realized cold atom Hubbard-Hofstadter lattice and will pave the way for observing possible higher order topological superfluidity with conventional $s$-wave superfluids/superconductors.
\end{abstract}

\pacs{}

\maketitle

\paragraph{Introduction:} $D$-dimensional topological insulators/superconductors (TI/TS) are characterized by a fully gapped bulk spectrum and stable gapless conducting states localized on $(D-1)$-dimensional  boundaries \cite{RMP_Kane,RMP_Shouchen}. Examples include the 3D strong TI with an odd number of gapless Dirac cones localized on 2D surfaces and the 1D spinless $p$-wave superconductors (SC) with zero-dimensional Majorana zero modes (MZMs) localized near the end points of the system. By contrast, the recently-introduced  so-called higher order TI/TS are gapped in the bulk as well as on the $(D-1)$-dimensional boundary, but have robust gapless topological modes  on $(D-2)$-dimensional ``edges'' defined on the boundary, e.g., corners in 2D systems and hinges in 3D systems. This idea has been used to explain the existence of protected low energy corner modes in 2D quantized eletric quadrupole insulators \cite{hughes_sci,hto_hughes,hto_fang,Neupert_sci,hto_Brouwer_1,hto_Brouwer_2,huber2018,Gaurav2018,Ronny2018} and the existence of 1D protected gapless hinge-modes in 3D crystals of bismuth \cite{Neupert2018_np}.

It has been recently proposed
\cite{2018_weakpairing,Wang2018_cornermode,Zhang2018_cornermode} that zero-dimensional Majorana corner modes (MCMs)
in 2D SC systems can be realized from a combination of 2D TI (QSHI) and unconventional (non-$s$-wave) superconductors. Excitations in these systems come in the form of MKPs, which are distinct from non-degenerate MZMs \cite{2001kitave,kane2008_TI,shoucheng_2010_TI,ktLaw_2009_AR,DasSarma_2010,DasSamar_2008,Zoller2007,DasSarma_2010_phasetrans,Fujimoto2009,Franz2011,Beenakker2009} and are protected by time reversal (TR) symmetry \cite{tri_sumanta,MKP_Chuanwei,MKP_40,MKP_41,MKP_42,MKP_43,MKP_44,MKP_45,MKP_46,MKP_47,MKP_48,MKP_49,MKP_50,MKP_51,MKP_52,MKP_53,MKP_54,MKP_55,MKP_56,MKP_58}. Unfortunately, MCMs proposed in the condensed matter systems \cite{weakpairing_22,weakpairing_23,weakpairing_24,2018_weakpairing,Wang2018_cornermode,Zhang2018_cornermode,wen_add,chen_add}  have not been realized to date.

In this paper, we propose ultracold atoms in optical lattices as a clean and  straightforward route to realize MCMs and higher order topological superfluidity with ordinary $s$-wave superfluids. 2D QSHI Hamiltonians have now been experimentally realized in cold atom systems on square optical lattices \cite{harperham2013}.  These systems are accurately modeled by a two-component Hofstadter model in a TR invariant scheme where the atoms experience opposite uniform magnetic fields for each of the two components \cite{harper1955,Hofstadter1976,Struck2013,harperham2013,harperham2013_opt,neutra_atoms,hofstadter_cold_atom_2012,Jotzu2014a}.
Furthermore, $s$-wave superfluidity can be induced with an attractive Hubbard interaction arising from a Feshbach resonance between the fermions \cite{Troyer2014_HHmodel,Iskin,Iskin2017_bcs,Hans_Feshbach,Chin_Feshbach,Salomon_Feshbach,Saenz_Feshbach}.
Specifically, we study a 2D TR invariant Hofstadter model, $H_0$, with an attractive Hubbard interaction, $H_I$:
$H=H_0+H_I.$
The model is characterized by an interaction-controlled phase transition between a QSHI and a superfluid (SF). Above a critical value of the attractive interaction, both the edge and the bulk have a non-zero superfluid order parameter due to BCS-like pairing. Since the edge spectrum is gapped, the 2D superfluid is topologically trivial, according to the conventional (lower order) bulk-boundary correspondence. We show, however, that the superfluid can host MKPs when a line soliton intersects the edges, changing the sign of the superfluid order parameter \cite{Wang2018_cornermode,Zhang2018_cornermode}. Dark solitons \cite{stringari2007darksoliton}, which have been successfully observed in Fermi gases \cite{Yefsah2013,Ku2014,Ku2016} using phase imprinting \cite{Condensate2009}, can arise as topological defects where the order parameter vanishes and the phase changes by $\pi$ \cite{Zhang2014soliton,Liu2015soliton_Majorana}. 
Intuitively, the edge states are gapped by the superfluid order parameter, which acts as a Dirac mass. At the intersection of the dark line soliton with the sample edges, the superfluid order parameter (hence the Dirac mass) changes sign, producing a pair of localized zero-dimensional MZMs protected by TR symmetry.  Tunneling into the soliton edges can be used to observe these MKPs \cite{Zhang2014soliton}. 
We emphasize that the uniform superfluid with no soliton is topologically trivial (in the conventional sense), with the appropriate $\mathcal{Z}_2$ invariant \cite{Ludwig2010} being trivially
zero because of the absence of gapless edge modes. Therefore, our work proposes the first cold atom-based realization of ($D-2$)-dimensional MZMs in what is a $D$-dimensional topologically trivial system in conventional sense and is thus an experimentally realizable higher order topological superfluid. 

\paragraph{Non-Interacting Model and Hofstadter Bands:} The Hofstadter model \cite{Hofstadter1976} describes non-interacting particles on a 2D lattice in the presence of a perpendicular magnetic field $\bm{B}=B \hat{z}$ given by the vector potential $\bm{A}=\left(0,B x,0\right)$. We consider a generalization of the original  model that includes a spin-dependent magnetic field:
\begin{eqnarray}
H_{0} &=& -\sum_{i,j,\sigma}\left[ t_x c^{\dagger}_{i,j,\sigma} c_{i+1,j,\sigma}+t_y^\sigma(i) c^{\dagger}_{i,j,\sigma} c_{i,j+1,\sigma} +h.c.\right] \nonumber \\
&+& \sum_{i,j,\sigma}\left(V_{i,j}-\mu\right) c^{\dagger}_{i,j,\sigma} c_{i,j,\sigma}, \label{eq:hof_ham}
\end{eqnarray}
where $(i,j)$ labels the sites of a square lattice with lattice constant $a$,  $c ^{\dagger}_{i,j, \sigma} \left( c_{i,j, \sigma} \right)$ creates $\left( \text{annihilates} \right) $ a particle at $\left(i,j \right)$ with spin $\sigma \equiv \{\uparrow, \downarrow\}$, $t_x=t$ and $|t_y^\sigma(i)|= t$ are nearest-neighbor hopping amplitudes along the $x$, $y$ directions respectively.  We set $t=1$ and consider i) periodic boundary conditions, ii) a cylindrical geometry (periodic in the $y$-direction and a finite width in the $x$ direction, $L_x= a N_x$), and iii) a rectangular geometry with $L_x= a N_x$, $L_y= a N_y$.  The chemical potential is $\mu$, while $V_{i,j}$ is a position-dependent confinement potential (See the Supplemental Material \cite{SM}).

In the presence of (spin-dependent) magnetic field, the hopping amplitude $t_y^\sigma(i)$ acquires a spin- and $x$-dependent phase factor $e^{i 2\pi \phi_{x_i,\sigma}}$, with  $\phi_{x_i,\sigma }=s_{\sigma} e B a x_i/ h$. Here,  $x_i =i a $ is position along the $x$-direction while $s_{\uparrow}= 1, s_{\downarrow}= -1$ correspond to opposite magnetic field orientations for the two spin components, which explicitly restores TR symmetry, in contrast with the original Hofstadter model.
We define the number of magnetic-flux quanta per unit cell as $\alpha=(B a^2)/\phi_0$, with $\phi_0=h/e$ the magnetic-flux quantum, such that $t_y^\sigma(\ell) =t e^{s_{\sigma} i 2\pi  \alpha \ell}$. For $\alpha = p/q$, with $p$ and $q$ primes, the single-particle energy spectrum is given by $q$ sub-bands $\epsilon_{\textbf{k}n}$, with $n=0,1,2,...,q-1$.  Here, we focus on the case $\alpha=1/3$.  We expect similar physics for other values of $\alpha$ that support QSHI phases.

In momentum space, $\mathbf{k}=(k_x,k_y)$, Eq.~(\ref{eq:hof_ham}) with $\alpha =0$ can be written as $H_0({\bm k})=-2t\sum_{k_x,k_y,\sigma} \left[ \cos\left( k_x\right ) + \cos\left( k_y\right )\right] $. The corresponding energy spectrum has a bandwidth of $8 t$ and the system is topologically trivial. To explore a topologically nontrivial regime, we consider $\alpha=1/3$ and use the Fourier transform $c_{i,j, \sigma}=N_0^{-1/2}\sum_{k} e^{i \mathbf{k r} } c_{k,\beta,\sigma}$, where $\mathbf{r}=(i,j)$, $N_0$ is the total number of lattice sites. The field-induced phase factors contained in $t_y^\sigma(\ell)$ give rise to  a new periodicity in the $x$-direction: $e^{i 2\pi s_{\sigma} \alpha \ell} =1,  e^{i 2\pi s_{\sigma} /3}, e^{i 4\pi s_{\sigma} /3}$ for $ \ell ~{\rm mod}~ 3 =0,1,2$,  respectively. We label the  non-equivalent sites in the $n^{\rm th}$ magnetic unit cell as $\beta=0,1,2$ such that $ x_\ell/a =\ell(n,\beta) =n q+\beta$. The corresponding first Brillouin zone is $k_x  \in [-\pi/q, \pi/q ]$, $ k_y \in [-\pi,\pi]$. After Fourier transforming, we can rewrite $H_0$  as
\begin{equation}
\begin{split}
H_0(\mathbf{k})=\sum_{\mathbf{k}\sigma} \psi^{\dagger}_{\mathbf{k},\sigma}\left(
\begin{array}{ccc} h_0  & e^{i k_y} & e^{-i k_y} \\
e^{-i k_y} & h_1 & e^{i k_y} \\
e^{i k_y} & e^{- i k_y} & h_2
\end{array}
\right)  \otimes I_2 \psi _{\mathbf{k},\sigma}
\end{split}\label{eq: k_space H_B}
\end{equation}
where $I_n$ is the $n \times n$ identity matrix,  $\mathbf{k}_{\beta}=(k_x- \beta 2\pi \alpha,k_y)$,
$\psi_{\mathbf{k},\sigma}=(c_{\mathbf{k}_1,\sigma},c_{\mathbf{k}_2,\sigma},c_{\mathbf{k}_3,\sigma})^T$,  $h_{\beta}=2 \operatorname{cos}\left( k_x -2\pi \beta \alpha \right)$, with $ \beta=0,1,2 $.
The corresponding band structure is characterized by $q=3$ spin-degenerate bands  with non-zero Berry curvature $\Omega_{\mathbf{k}}^\sigma$ and non-zero spin-dependent Chern number.  Although the total Chern number of a fully-filled band is zero due to TR symmetry \cite{chern2015}, the corresponding $\mathcal{Z}_2$ invariant reveals a topological-nontrivial QSHI phase. \\
\begin{figure}
\begin{center}
\includegraphics[width=0.48\textwidth]{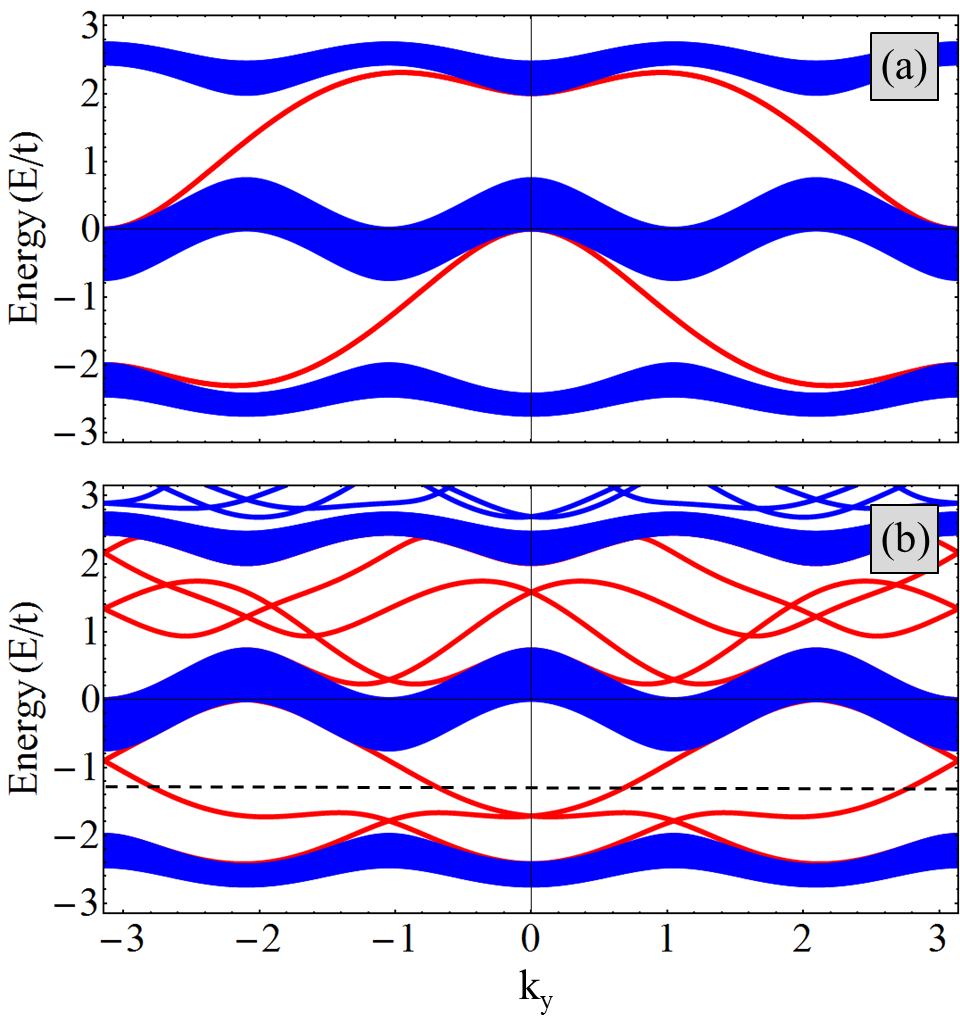}
\end{center}
\vspace{-4mm}
\caption{Band structure of the non-interacting two-component Hofstadter model, $H_0$, with $\alpha = 1/3$ and periodic boundary conditions in the y-direction. (a) System with $N_x=92$ and hard confinement in the x-direction. (b) System with $N_x=92$ and soft (Gaussian) confinement \cite{SM}. The bulk states are shown in blue and the red curves represent the gapless edge modes. \vspace{-1mm}}\label{fig:band_structure}
\end{figure}                     
 \indent
 The characteristic edge modes can be obtained using a cylindrical geometry with periodic boundary conditions in the $y$-direction.
The corresponding band structure for a system with both  hard and soft confinement \cite{SM} is shown in Fig.~\ref{fig:band_structure}. The red lines indicate the (confinement-dependent) gapless edge states, while the (dense) blue lines correspond to the bulk spectrum.
When the chemical potential intersects the red lines, e.g., at  $\pm \mathbf{k}_{\sigma} $, the system supports a pair of gapless edge states $\left( \mathbf{k}_{\uparrow}, -\mathbf{k}_{\downarrow}\right)$ located along one of the edges and another pair $ \left( -\mathbf{k}_{\uparrow}, \mathbf{k}_{\downarrow}\right)$ located on the other edge. Consequently, if $\mu$ lies within a bulk gap, the system is in a topological QSHI phase with pairs of counter-propagating gapless modes located along the edges.

\vspace{1mm}
\noindent
\paragraph{Attractive Interactions:}
Next, we introduce an attractive interaction described in real space by the  Hubbard term
\begin{equation}
\setlength\abovedisplayskip{6pt}
\setlength\belowdisplayskip{4pt}
H_I =- U \sum_{i,j} n_{i,j,\uparrow} n_{i,j,\downarrow},
\end{equation}
where $U>0$ is the magnitude of the on-site attraction. In cold atom systems, the interaction can derive from an attractive Feshbach resonance \cite{Hans_Feshbach,Saenz_Feshbach}.
We study the effect of this attractive interaction at the mean-field level using a BCS-like approximation. In $k$-space, we have
\begin{equation}
\begin{split}
H_I \rightarrow \sum_{\mathbf{k},\beta} \left (\Delta^{\dagger} c_{-\mathbf{k},\beta \downarrow}  c_{\mathbf{k},\beta \uparrow}+ \Delta c^{\dagger}_{\mathbf{k},\beta \uparrow} c^{\dagger}_{-\mathbf{k},\beta \downarrow} \right)+\frac{3N_0}{U} |\Delta|^2,
\end{split} \label{eq:mf_order}
\end{equation}
where we have introduced a uniform \cite{Iskin2015_order} order parameter $\Delta =-(U/N_0) \sum_{\mathbf{k}} \left< c_{-\mathbf{k},\beta \downarrow}  c_{\mathbf{k},\beta \uparrow} \right>$, with $\left< ... \right>$ indicating the thermal average. 
At this mean-field level, the total  Hamiltonian becomes
\begin{equation}
\begin{split}
H_{MF}=\sum_{\mathbf{k}} \Psi^{\dagger}_{\mathbf{k}} \left(
\begin{array}{cc} h_{B}(\mathbf{k})-\mu &  \Delta \mathbf{\cdot} I_3 \\
\Delta^{\dagger} \mathbf{\cdot} I_3 &  - h^{*}_{B}(-\mathbf{k})+\mu
\end{array}
\right) \Psi _{\mathbf{k}} +\mathcal{E}
\end{split}\label{eq: MF_H}
\end{equation}
where $\Psi_{\mathbf{k}}=(\psi_{\mathbf{k}, \uparrow},\psi^{\dagger}_{-\mathbf{k}, \downarrow}) ^{T}$, $h_B$ is the matrix in Eq.~(\ref{eq: k_space H_B}), and $ \mathcal{E}=- \sum_{\mathbf{k}} \left ( 3\mu-3|\Delta|^2/U +\mathbf{Tr} E_{-\mathbf{k},\downarrow}
\right)$ is an energy offset.  
We solve this model using a self-consistent BCS-like formalism outlined in the Supplemental Material \cite{SM}.

\begin{figure}
	\begin{center}
		\includegraphics[width=0.48\textwidth]{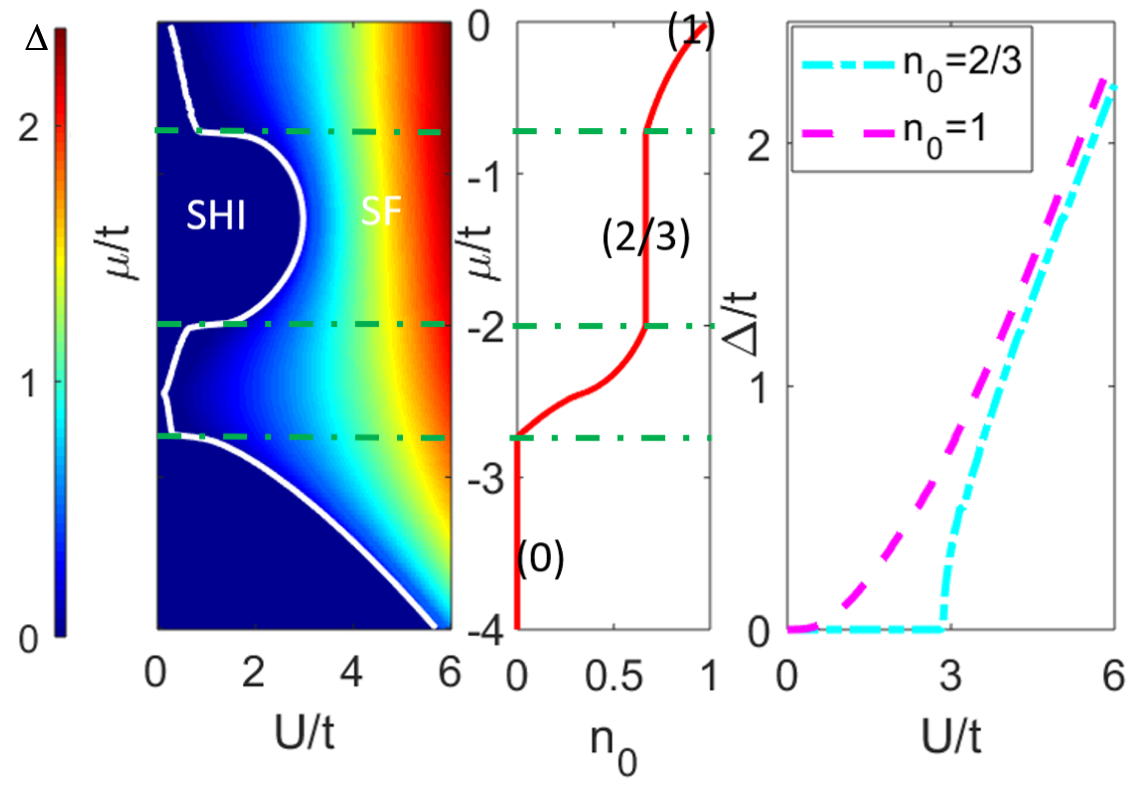}
	\end{center}
	\vspace{-6mm}
	\caption{(Color online)
\emph{Left}: The mean-field phase diagram obtained by plotting the self-consistent value of the pairing order parameter $\Delta$ for $k_B T=10^{-4}t$. The dashed-white line indicates the phase boundary.  \emph{Center}: Chemical potential as function of the filling factor for $\Delta =0$.  \emph{Right}: Mean-field $\Delta$ as function of the interaction strength for two different filling factors.  The $n_0= 1$ line shows $\Delta\neq 0$ (i.e. superfluid phase) all the way to $U \sim 0$,  while for $n_0= 2/3$ one needs $U \sim 3t$ to enter the superfluid phase.. The green dotted lines mark the band edges of the bulk spectrum in Fig.~\ref{fig:band_structure}.}\label{fig:phase diag}
\end{figure}


The mean-field phase diagram corresponding to Eq.~(\ref{eq: MF_H}) is shown in Fig.~\ref{fig:phase diag}. When the chemical potential lies within the bulk gap, the self-consistent value of the s-wave pairing becomes non-zero only above a finite interaction strength $U_c(\mu)$. For $U<U_c(\mu)$ the system is in a QSHI phase with $\Delta=0$, while $U> U_c(\mu)$ corresponds to the superfluid phase ($\Delta\neq 0$).
Note that for $\mu \in [-2t,-0.7t] $, the phase transition from a QSHI state with filling factor $n_0  = 2/3 $ to the SF state occurs at a critical interaction on the order of $3t$. On the other hand,  at  half filling ($n_0=1$) $\Delta\neq 0$ for any finite $U$ and the system is in a SF phase. Below, we will show the SF phase supports MKPs in the presence of a line soliton, when the order parameter changes sign.

\begin{figure}
	\begin{center}
		\includegraphics[width=0.48\textwidth]{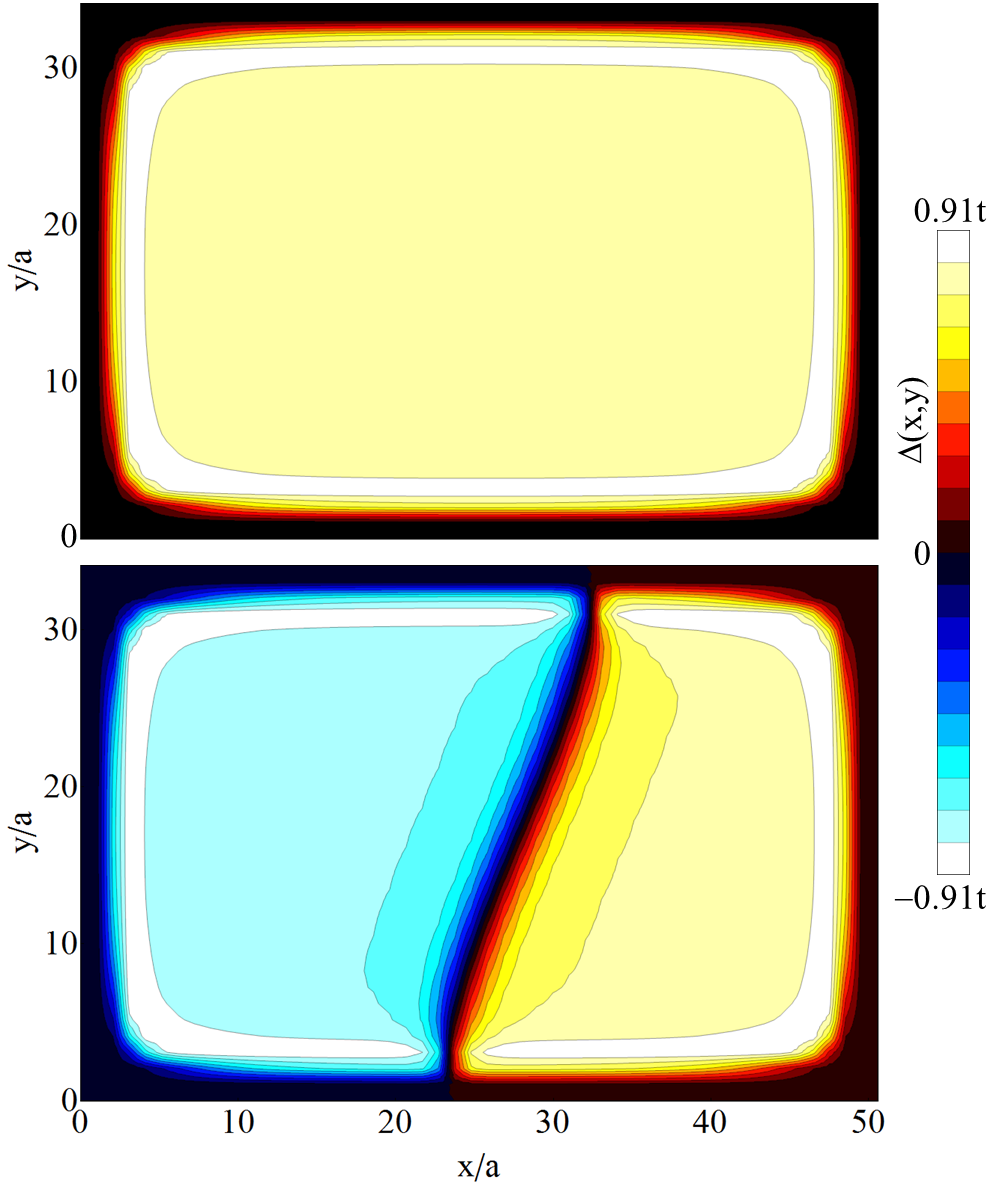}
	\end{center}
	\vspace{-4mm}
	\caption{Position-dependent pairing potential $\Delta(x,y)$ for a strongly interacting system with $U=3.5t$, i.e., in the SF phase. The pairing potential is obtained as the self-consistent solution of the mean-field equations (S7-S10) for a finite system with $N_x\times N_y = 50\times 34$ and soft confinement \cite{SM} at finite temperature, $k_BT=0.01t$. The total number of particles is fixed, $N = 800$. {\em Top}: Self-consistent solution with constant phase. The (self-consistent) chemical potential is $\mu=-1.250t$. {\em Bottom}: Self-consistent solution with a line soliton. The chemical potential is $\mu=-1.248t$. Note $\Delta(x,y)$ is nonzero in the bulk -- consistent with  phase diagram in  Fig.~\ref{fig:phase diag}(b) -- as well as on the boundary of the system, except along the line soliton. }\label{fig: soliton density strong}
\end{figure}
\paragraph{Soliton-induced Majorana Zero-energy Modes:}  Next, we show in the presence of a line soliton, MKPs emerge at the corners defined by the intersection of the soliton with the edge of the system, which is in a TR symmetric SF phase. In the presence of a dark soliton, the order parameter changes sign, vanishing along a node line. To study the impact of the soliton,  we construct the BdG equations in real space and solve them self-consistently \cite{SM}.  We choose the initial value of the order parameter to be used in the self-consistent scheme as:
$\widetilde{\Delta}_{i, j} = \Delta_{i,j}\tanh [(i-28+5\cos[(j-1)\pi/(N_y-1)])/\xi]$, where $\Delta_{i,j}$ is a constant phase self-consistent solution (i.e. obtained without the soliton) and  $\xi=2.5$.  We then solve the BdG equations for $N_x \times  \ N_y =50\times 34$ sites and $n_0 =2/3$.

\begin{figure}
\begin{center}
\includegraphics[width=0.48\textwidth]{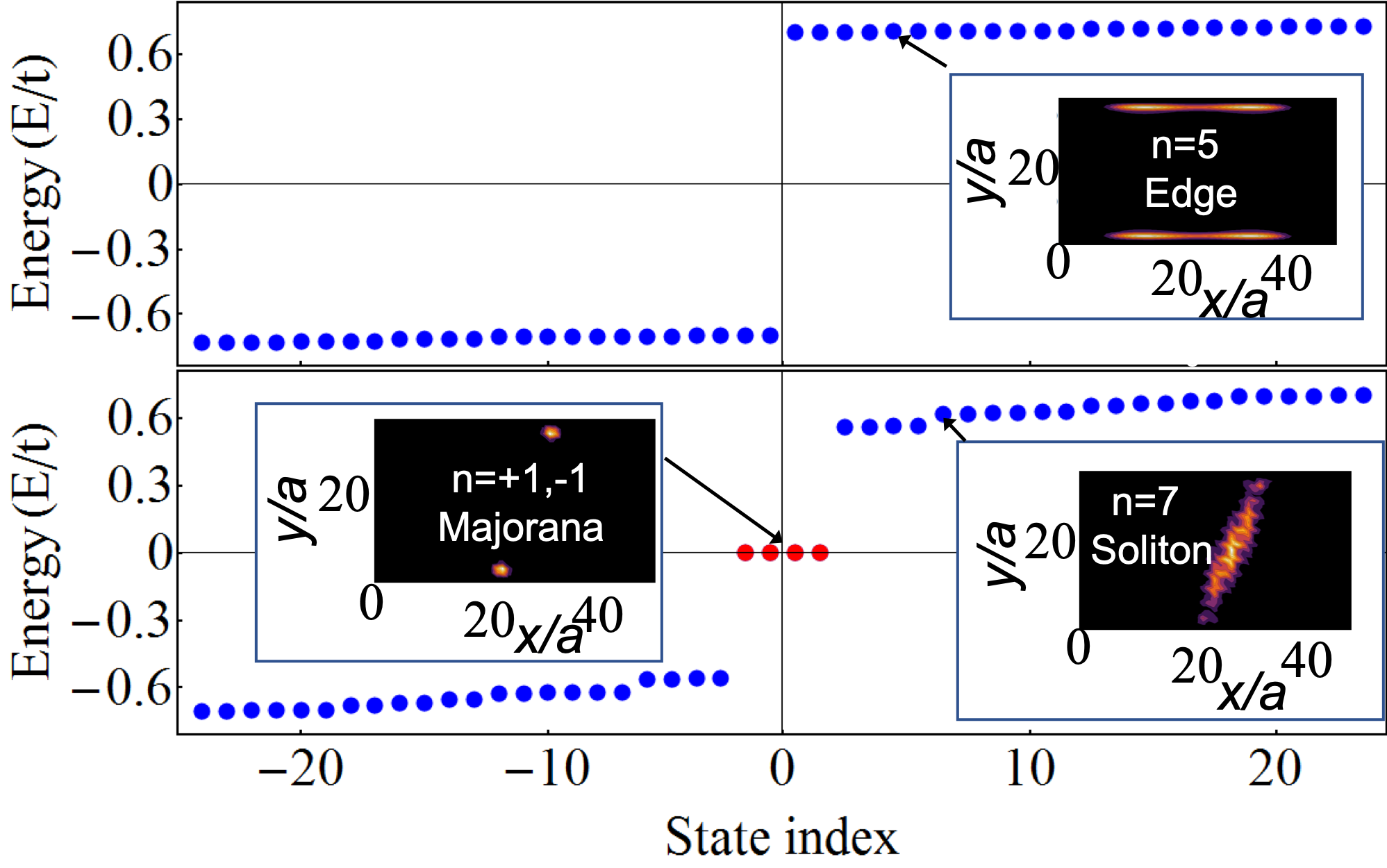}
\end{center}
\vspace{-4mm}
\caption{\emph{Top}: Low-energy spectrum of the Hofstadter-Hubbard model  with strong interaction ($U=3.5t$) within the mean-field approximation for a system with constant phase pairing potential and parameters corresponding to Fig.~\ref{fig: soliton density strong} (top panel). \emph{Bottom}: The same but for parameters corresponding to the bottom panel of Fig.~\ref{fig: soliton density strong}. Note in the presence of a line soliton the system hosts two pairs of zero-energy Majorana bound states (red dots). The insets plot the wave functions of the states marked by arrows. \vspace{-4mm}  }
\label{fig:Spectrum}
\end{figure}

 In Fig.~\ref{fig: soliton density strong} we show the self-consistent solution for $\Delta(x,y)$ for a system with $U = 3.5 t$ without (top panel) and with a line soliton (bottom panel). Note $U>U_c(\mu)$, so that, without the soliton, the system  is in a SF phase with a non-vanishing order parameter both in the bulk and on the edge.  The bottom panel shows the soliton changes the sign of the order parameter, as expected.  Note for $U<U_c(\mu)$  the order parameter vanishes in the bulk, but remains finite on the edge, where it changes sign in the presence of a soliton \cite{SM}.

The low-energy spectra corresponding to the self-consistent solutions in Fig.~\ref{fig: soliton density strong} are shown in Fig.~\ref{fig:Spectrum}.
The top panel (no soliton) is characterized by a finite quasiparticle gap and low-energy states located along the edges of the system (see inset).
The bottom panel, corresponding to a system with a soliton,  has four zero-energy states (red circles) representing the MKPs.
As shown in the inset, the corresponding wave functions are localized at the intersection of the line soliton with the edges of the system.

Our results show MKPs can be induced at soliton edges in a conventional s-wave SF. We have checked the line soliton and the corresponding MKPs are robust against small perturbations (e.g., thermal fluctuations and on-site disorder, Ref.~\cite{SM}) and are thus topologically robust.  
\paragraph{Implementation:} To implement $H_0$ we envision an experimental setup similar to Ref.~\cite{harperham2013} since this scheme does not rely on the internal atomic structure.  We consider a 3D cubic optical lattice were confinement along $z$ separates the system into parallel $x-y$ planes.  The 2D Hubbard model then approximates the dynamics of $^{40}$K or $^6$Li placed with one atom per site in a deep optical lattice with uniform hopping $t$ if we equally populate two Zeeman levels with opposite magnetic moments.  A magnetic field gradient along the $y$-direction creates a splitting (much larger than $t$) between opposite spins in neighboring sites.  In addition to the primary lattice beams, a pair of running-wave beams are applied parallel to the $x-y$ bonds of the square lattice to dynamically restore resonant tunneling assuming the running-wave lattice depth is much smaller than the spin splitting.  This setup induces the complex spin-dependent phase in Eq.~(\ref{eq:hof_ham}) in a rotating wave approximation. 

To implement $H_I$ we require an attractive Feshbach resonance.  For magnetic Feshbach resonances, typical magnetic field gradients ($\sim 10mG/\mu \text{m}$) leave the attractive interaction spatially uniform since common resonances occur at relatively high fields ($\sim 400-700 G$) and can be broad, as in, e.g., $^6$Li.  It is also safe to assume close proximity to the Feshbach resonance does not lead to strong heating and loss \cite{Williams2013} since the Raman coupling \cite{harperham2013} between the same hyperfine states (and neighboring lattice sites) does not induce any new three-body loss channel.  

Tuning the chemical potential near zero (Fig.~\ref{fig:Spectrum}) allows observation of MZMs.  Spatially resolved radio-frequency spectroscopy and probing of the density profile have been proposed as an experimental approach to detect these MZMs \cite{Liu2015soliton_Majorana,Liu2013probing}.  The soliton-induced MZMs can be manipulated by controlling the spatial location of the soliton excitation, which may be beneficial for topological braiding \cite{Nayak2008,Zhang2007} of MZMs.
\paragraph{Discussion and Conclusion:}
The essential physics for the creation of MKPs and higher order topological superfluidity in the current system is similar to the proposals for higher order topological superconductors in solid state systems. 
In both cases, the non-SC ``normal'' system is a 2D QSHI. 
This system has counter-propagating Kramer’s pairs of gapless edge states (Fig.~\ref{fig:band_structure}), which can support spin-singlet superconductivity. Furthermore, in both systems introducing 
superconductivity (by proximity effect in solid state systems and interaction-induced, via Feshbach resonance, in the present work) gaps out the edge modes, which signals the system is a topologically trivial superconductor/superfluid (because the edge modes are gapped). However, whenever the superconducting gap changes sign (thus goes through zero) at a point along the edge, a Kramer’s pair of localized MZMs are nucleated by the Jackiew-Rebbi mechanism, which is common to both the solid state proposals and the present work (a Kramer’s pair of zero modes is nucleated because
the system is time reversal invariant).

The key difference between the solid state case and the current set-up is that in the former system the change of sign of the superconducting gap is proposed to be realized by proximity effect with an unconventional superconductor (such as $d$-wave 
or $s_{\pm}$wave, which change sign in momentum space), while in our work the change of sign of the superconducting gap is due to
a soliton in the $s$-wave superfluid. The other significant difference between the two proposals is that, while the proximity effect of unconventional $d$ or $s_{\pm}$wave superconductivity on QSHI in solid state systems has not yet been demonstrated experimentally (and is probably going to be hard) the main ingredients of the same physics within our proposal, namely, the two-component Hofstadter model (thus a QSHI, ~\cite{harper1955,Hofstadter1976,Struck2013,harperham2013,harperham2013_opt,neutra_atoms,hofstadter_cold_atom_2012,Jotzu2014a}), on-site attractive interactions and non-zero SC pair potential~\cite{Troyer2014_HHmodel,Iskin,Iskin2017_bcs,Hans_Feshbach,Chin_Feshbach,Salomon_Feshbach,Saenz_Feshbach}, and creation of dark solitons~\cite{stringari2007darksoliton,Yefsah2013,Ku2014,Ku2016}, have all been individually realized in the cold atom systems.

C. Zeng, S.T. and V.W.S. acknowledge support from ARO Grant No. W911NF-16-1-0182. T.D.S. was supported by NSF Grant
No. DMR-1414683. V.W.S. acknowledges support from AFOSR (FA9550-18-1-0505). C. Zhang is supported by NSF (PHY-1505496, PHY-1806227), ARO (W911NF-17-1-0128), and AFOSR (FA9550-16-1-0387).

\bibliography{my}

\end{document}